# Enhanced optoelectronic performance and photogating effect in quasi-one-dimensional BiSeI wires

H. J. Hu,[1,2,a)] W. L. Zhen,[1,a)] S. R. Weng,[1] Y. D. Li,[1] R. Niu,[1] Z. L. Yue,[1] F. Xu,[1] L. Pi,[1,2,b)] C. J. Zhang,[1,3,b)] and W. K. Zhu[1,b)]

[1]High Magnetic Field Laboratory, Chinese Academy of Sciences, Hefei 230031, China

[2]University of Science and Technology of China, Hefei 230026, China

[3]Institutes of Physical Science and Information Technology, Anhui University, Hefei 230601, China

a)H.J.H. and W.L.Z. contributed equally to this work.

b)Authors to whom correspondence should be addressed: wkzhu@hmfl.ac.cn, pili@ustc.edu.cn and zhangcj@hmfl.ac.cn

## ABSTRACT

Quasi-one-dimensional (quasi-1D) materials are a newly arising topic in low-dimensional researches. As a result of reduced dimensionality and enhanced anisotropy, the quasi-1D structure gives rise to novel properties and promising applications such as photodetectors. However, it remains an open question whether performance crossover will occur when the channel material is downsized. Here we report on the fabrication and testing of photodetectors based on exfoliated quasi-1D BiSeI thin wires. Compared with the device on bulk crystal, a significantly enhanced photoresponse is observed, which is manifested by a series of performance parameters, including ultrahigh responsivity ($7 \times 10^4$ A W$^{-1}$), specific detectivity ($2.5 \times 10^{14}$ Jones) and external quantum efficiency ($1.8 \times 10^7$%) when $V_{ds} = 3$ V, $\lambda = 515$ nm and $P = 0.01$ mW cm$^{-2}$. The conventional photoconductive effect is unlikely to account for such a superior photoresponse, which is ultimately understood in terms of the increased specific surface area and the photogating effect caused by trapping states. This work provides a perspective for the modulation of optoelectronic properties and performance in quasi-1D materials.

Low-dimensional materials have attracted considerable attention for their intriguing mechanical, electrical, optical, and physiochemical properties originating from the reduced dimensionality and quantum confinement effects,[1] and also for the compatibility with device fabrication and integration and other useful functional applications.[2, 3] As a large family, two-dimensional (2D) materials and devices have been intensively studied and their various applications in the fields of physics,[4] chemistry,[5] materials[6] and biology[7] have been explored. Quasi-one-dimensional (quasi-1D) materials are a newly arising topic in low-dimensional researches,[8] although many of them have been discovered or synthesized for a long time.[9, 10] Compared with three-dimensional (3D) and 2D materials, the quasi-1D structure is unique as a result of the enhanced anisotropy, which gives rise to novel properties such as electronic instability,[11] prominent surficial effect,[12] high current-carrying capability,[13, 14] charge density wave,[14, 15] superconductivity,[16] etc. Quasi-1D materials usually consist of long atom chains. The interactions between chains are relatively weak chemical bonds or van-der-Waals (vdW) force, which makes the growth commonly in one direction and forms needle- or wire-shaped crystals. Moreover, the crystals can be easily peeled off into thin wires.

For quasi-1D semiconductors, the optoelectronic properties play a fundamental role in photoresponsive devices like photodetectors,[17-19] chemical sensors,[20] light switch[21] and artificial neural networks.[22] BiSeI is a representative quasi-1D semiconductor that shows a significant photoresponse. Its crystal structure is composed of double chains $[(BiSeI)_\infty]_2$ along the *b* axis, where Bi and Se atoms are connected by covalent bonds and I ions have ionic bonds with a covalent binding bridge.[23] BiSeI is an indirect bandgap semiconductor with a bandgap of ~ 1.32 eV, which has been confirmed by the first-principles calculations[24] and reflectivity measurement.[25] In addition, BiSeI has been studied as a promising photovoltaic material,[26] manifesting the application potential in optoelectronics. Indeed, the photodetector constructed on the bulk crystal shows an obvious photoresponse under visible light.[24] Considering that the reduced size and dimensionality of quasi-1D materials usually cause a crossover in properties, the reduced size of BiSeI crystal is expected to induce a significant change in optoelectronic performance. Moreover, a non-stoichiometric effect and iodine vacancies are observed in the bulk crystals. It remains an open question how the non-stoichiometry would affect and modulate the photoresponse of BiSeI thin-wire devices.

In this work, the photodetectors are manufactured based on the exfoliated BiSeI thin wires of different thicknesses. Compared with the bulk device, the photoresponse of thin-wire devices is significantly enhanced, and the performance parameters are at least 3 orders of magnitude higher. The conventional photoconductive effect is unlikely to account for the main mechanism, which is eventually understood in terms of the increased specific surface area and the photogating effect arising from trapping states. This work shows a possible way to modulate the optoelectronic properties and performance in quasi-1D materials.

High-quality single crystals of BiSeI were grown by the physical vapor transport (PVT) method.[24] Single crystal X-ray diffraction (XRD) measurement was performed on a Rigaku Miniflex X-ray diffractometer using Cu Kα radiation. Raman spectra were collected on a Horiba JobinYvon T6400 spectrometer with a 532 nm laser. Using Nitto SPV-224R tape, the crystals were peeled off into thin wires that were further transferred to the $SiO_2$/Si substrates. The electrodes were prepared by ultraviolet lithography and thermal evaporation coating process. The morphology was characterized using a scanning electron microscope (SEM) and an Olympus BX53M optical microscope. The chemical composition was characterized on an Oxford energy dispersive spectroscope (EDS). The electrical measurements were taken on a probe station combined with a Keithley 2636B system source meter. Lasers of different wavelengths (405 nm, 450 nm, 488 nm and 515 nm) were used in the optoelectronic measurement.

The quasi-1D structure of BiSeI makes the result of mechanical exfoliation different from typical 2D materials. As illustrated in Fig. 1(a), the needle-shaped crystals can be peeled off with tape and transferred to the substrate. Only long and straight thin wires are found to remain on the substrate. The morphology of the wires is characterized using SEM (Figs. 1(b)-1(d) and Fig. S3 in the supplementary material), with thicknesses ranging from 70 nm to a few microns. As the thickness decreases, the wire becomes rounder. Thick wires have poor roundness. In the present study, we focus on wires that are about 1 $\mu$m thick or thinner. As shown in Figs. 1(b)-1(d), three devices are fabricated based on wires of different thicknesses, namely 145 nm (labeled D1), 581 nm (D2) and 1.18 $\mu$m (D3). Figures 1(b) and 1(c) are SEM images of D1 and D2, and Fig. 1(d) is an optical image of D3. It can be seen that the channel length is much larger than the channel width. The contact between the electrodes and the wires exists not only on the top but also on the side of the wires, because the electrode metal is only 100 nm thick. The results of structural and chemical characterizations are presented in the supplementary material. It is worth noting that there are some Se and I vacancies in the crystal that cannot be ignored. Their concentrations are calculated to be 5.4% and 10.5%, respectively. The vacancies will induce self-doping effects, which are expected to affect the properties of the channel material used in the devices. Other types of defects, such as antisite and interstitial defects, need to be characterized in future work.

  Basic electrical measurements are first performed to study the semiconductor properties. As shown in Fig. 2(a), the current-voltage ($I_{ds}$-$V_{ds}$) characteristic of D1 passes through the origin and is almost linear within $\pm 3$ V, indicating that the contact is an ohmic contact. The conductivity is calculated as 0.21 S m$^{-1}$. The $I_{ds}$-$V_{ds}$ curves of D2 and D3 are included in the supplementary material. Figure 2(b) shows the transfer curve ($I_{ds}$-$V_{gs}$) of D1 in the range of $\pm 80$ V. From the semi-logarithmic plot (right axis), although the $I_{ds}$ at -80 V is as low as 10 pA and increases rapidly with the increase of $V_{gs}$, it does not reach a clear reverse cutoff state, which suggests that the thin wire used in D1 is a heavily doped n-type semiconductor. The ratio of $I_{ds}$(80 V)/$I_{ds}$(-80 V) exceeds $10^3$, which is a comparably prominent value.[27] Using the formula $N_d = \frac{dV_{gs}}{dI_{ds}} \times \frac{I_{ds} C_i}{ed}$, where $e$ is the elementary charge, $d$ is the channel thickness, and $C_\text{i}$ is the capacitance per unit area (11.6 nF cm$^{-2}$ for the 300 nm SiO$_2$ layer),[28] the carrier concentration $N_\text{d}$ is calculated to be $8.9 \times 10^{17}$ cm$^{-3}$ at zero gate. This value is relatively large compared with typical n-Si and n-GaAs,[29] consistent with the heavily doped nature.

The optoelectronic tests are further carried out under laser illumination of various wavelengths and tuned to different powers. Figure 2(c) is a schematic illustration of optoelectronic measurement. The previous study has shown that BiSeI has a significant response to the entire visible-light spectrum, with the largest response to green light.[24] Figures 2(d)-2(f) show the $I_\text{ds}$-$V_\text{ds}$ curves obtained in the dark and under a 515 nm laser for D1, D2 and D3, respectively. The data obtained with the lasers of other wavelengths are presented in the supplementary material. A remarkable photoresponse is observed in all devices. The curves of D1 maintain good linearity for various laser power densities ($P$). Benefiting from the low dark current $I_\text{dark}$, the photocurrent, i.e., $I_\text{ph} = I - I_\text{dark}$, persistently increases as $P$ increases.

Although the photoresponse of D2 and D3 is also remarkable, their line shapes are slightly different. For the $I_\text{ds}$-$V_\text{ds}$ curves taken without illumination, the linearity cannot hold beyond $\pm 1$ V, and the symmetry is not as good as D1. This suggests existence of contact resistance due to the Schottky barrier. The reason why the Schottky barrier becomes non-negligible may be due to the increase in sample size. Considering that the electrode metal is only 100 nm thick and that the thickness of the wires in D2 and D3 exceeds 500 nm, the contact is probably not good enough to be an ohmic one. The conductivity at zero bias is calculated as 0.03 S m$^{-1}$ for D2 and 0.02 S m$^{-1}$ for D3. This confirms the existence of contact resistance. However, the contacts in D2 and D3 are acceptable for optoelectronic research, because the barrier effect on photocurrent is not obvious, and the dominant mechanism is photogating that will be discussed below.

The optoelectronic performance of the devices can be evaluated by several parameters such as response time ($\tau$), responsivity ($R$), specific detectivity ($D^*$) and external quantum efficiency (EQE, $\eta$). Figure 3(a) shows the $I_{ds}$ of D1 as a function of time taken at $V_{ds}$ = 3 V, $\lambda$ = 488 nm and $P$ = 21.2 mW cm$^{-2}$. The laser excitation is repeatedly turned on and off with a period of 10 s. It can be seen that the ON and OFF states are stable and well defined, showing good stability and durability of the device. The response time is calculated using the 10%-90% method. That is, the rising time ($\tau_{rise}$) and recovery time ($\tau_{decay}$) are defined as the time between 10% and 90% of the maximum photocurrent at the rising and recovery edges, respectively. The calculated response time is $\tau_{rise}$ = 0.6 s and $\tau_{decay}$ = 1 s [Fig. 3(b)]. Compared with the bulk device, the response process is slightly slower, which may be due to the existence of defects, namely the chemical nonstoichiometry suggested by the EDS data. Since there are more vacancies in thin wires than bulk crystals, it is understandable that the response time becomes longer.

Responsivity is the ratio of photocurrent to effective incident power, which represents the conversion efficiency from incident light signal to electrical signal.[30] The responsivity is defined as $R = \frac{I_{ph}}{PS}$, where $S$ is the illuminated device area. Figure 3(c) shows the power density dependence of $R$ calculated from the measurements taken at $V_{ds}$ = 3 V and $\lambda$ = 515 nm for D1, D2 and D3. It can be seen that the responsivity drops rapidly with the increase of $P$ and gradually stabilizes upon further increasing $P$. Under illumination, electrons in the valence band absorb photon energy and hop to the conduction band, generating electron-hole pairs and photocurrent. As $P$ increases, the absorption of light is quickly saturated, and the recombination of electrons and holes is enhanced, so the responsivity falls.[31] The maximum responsivity of D1 is 7 × 10$^4$ A W$^{-1}$ when $V_{ds}$ = 3 V and $P$ = 0.01 mW cm$^{-2}$. The responsivity at 0.1 V bias, which reads 1.89 × 10$^3$ A W$^{-1}$, is about 3 orders of magnitude higher than that of the bulk device ($R$ = 3.2 A W$^{-1}$ at $V_{ds}$ = 0.1 V, $\lambda$ = 635 nm and $P$ = 0.23 mW cm$^{-2}$).[24]

Moreover, the devices reveal an evolution as the thickness changes. To present the evolution in a clearer and more systematic manner, more devices are included and $R$ is plotted as a function of channel width (roughly in line with thickness) at different power densities. As shown in Fig. S8 in the supplementary material, for the lowest $P = 0.01$ mW/cm$^2$, $R$ increases monotonically with the decreasing width. As $P$ increases, this tendency is gradually weakened, especially for thick-wire devices. For the 1.18 µm device, it appears that $R$ begins to rise with the increasing width. This non-monotonic evolution exemplifies two factors with opposite size effects. Here we propose a possible scenario. Thick samples with more layers probably absorb more light, which will increase $R$. This factor is labeled I. The second factor (II) is specific surface area. For thinner wires, a larger specific surface area leads to a higher proportion of photogenerated current in the total current. This is clearly confirmed in Figs. 2(d)-2(f). As the thickness decreases, factor I is diminished and factor II is enhanced. From 1.18 µm to 581 nm, factor I outperforms factor II and $R$ decreases. From 581 nm to 101 nm, factor II outperforms factor I and $R$ sharply increases. It is easy to understand that factor I dominates at the bulk extreme, while factor II dominates at the nanoscale extreme. For bulk or thick samples, the change in specific surface area can be ignored, so factor I is dominant. For nanoscale samples, the light absorption is stable, but the specific surface area varies significantly, so factor II becomes dominant. For low laser power, factor II is more important because the light absorption rate is higher, which means similar absorption for different devices.

Factors related to surface conditions may explain the dependence of $R$ on specific surface area. On the one hand, the difference between the surface state (caused by surface defects or simply by the air-semiconductor interface) and the inner state of the nanowires can induce band bending and a built-in potential. This factor has the same effect as factor I.[32] On the other hand, the adsorption of gas molecules in the atmosphere may cause surface trapped states, especially for highly electronegative oxygen.[21, 33, 34] This oxygen-assisted recombination mechanism is further enhanced in thinner nanowires due to the increased surface-to-volume ratio and easily outweighs other factors.[34] This is a possible physical origin behind the specific surface area description, which requires further verification.

The specific detectivity, defined as $D^* = R\sqrt{\frac{S}{2eI_{\text{dark}}}}$,[35] characterizes the capability to detect the lowest light signal and is designed to compare photodetectors with different geometries. The EQE is defined as the number of electrons excited by each incident photon, expressed by $\eta = \frac{hc}{e\lambda} R \times 100\%$, where $h$ is Planck's constant and $c$ is the speed of light in vacuum.[24] From Figs. 3(d) and 3(e), similar rules are found again. That is, $D^*$ and $\eta$ decrease as $P$ increases, and their evolution with the wire thickness is almost the same as responsivity. The $D^*$ is up to $2.5 \times 10^{14}$ Jones and EQE is up to $1.8 \times 10^7\%$. Compared with the values obtained on bulk crystal ($D^* = 7 \times 10^{10}$ Jones and $\eta = 622\%$ under the same conditions as above),[24] the increase is also remarkable.

Figure 3(f) shows the photocurrent of all devices as a function of $P$. The curves are not linear but quickly approach saturation. They are fitted to the power law, i.e., $I_{\text{ph}} \propto P^\alpha$, where $\alpha$ is the power exponent. $\alpha = 1$ corresponds to the ideal condition of the photoconductive effect.[31] The fitting results give $\alpha = 0.31$ for D1 and $\alpha = 0.39$ for D2 and D3. Considering that the $\alpha$ of the bulk device is 0.42,[24] these values are far away from 1, indicating that the mechanism in all BiSeI devices deviates from the photoconductive model and nanoscale devices make the deviation more prominent. More analysis results are included in the supplementary material.

In order to investigate the underlying mechanism, the optoelectronic measurements modulated by the back gate are further performed. Figure 4(a) shows the $I_{\text{ds}}$-$V_{\text{ds}}$ curves of D4 obtained in the dark and under a 405 nm laser. The curves pass through the origin and show good linearity, indicating that the contact is an ohmic contact. A significant photoresponse is also observed for this device. Note that the jumps appearing around 10 nA are due to a change in the current measurement range, which is not intrinsic. Under the same condition of illumination, the transfer curves are measured and plotted in Fig. 4(b). Different from the other devices, the transfer curve taken in the dark shows a definite cutoff state below $V_{\text{gs}} = -65$ V ($V_{\text{th}}$, threshold voltage), suggesting that the stoichiometry condition in this wire is slightly different. The ON/OFF ratio is determined to be $2 \times 10^3$. Once the laser illumination is turned on (even for a small power), a significant photocurrent occurs and the cutoff state disappears. Upon further increasing $P$, the photocurrent gradually approaches saturation. When the back gate sweeps from -80 V to 80 V, the dependence of $I_{ph}$ on $P$ seems to changes greatly. To check the evolution of the mechanism, the $I_{ph}$ and $R$ are calculated for different $V_{gs}$. Figures 4(c) and 4(d) show $I_{ph}$ (left) and $R$ (right) as a function of $P$ for $V_{gs}$ = -80 V and $V_{gs}$ = 80 V, respectively. The $I_{ph}$-$P$ data are fitted to the power law and the power exponent $\alpha$ is obtained. The values of $\alpha$ for different $V_{gs}$ are listed in Fig. 4(e). Surprisingly, the data points distribute in a narrow range from 0.18 to 0.26. All the values are far away from 1, indicating that the photoconductive effect is unlikely to account for the main photocurrent-generation mechanism in BiSeI devices.

Generally, there are three possible main mechanisms for the photoresponse, i.e., thermal mechanism, photoconductive effect and photogating.[36] The thermal mechanism can be ignored here, because the laser illumination is uniform and covers the entire channel. The vertical shift of the transfer curve shown in Fig. 4(b) can be attributed to the photoconductive effect. The photocurrent generated from the photoconductive effect causes an increase in $I_{ds}$ through extra photogenerated carriers. In addition, the disappearance of the cutoff state and $V_{th}$ under illumination is a signature of photogating. The shift of $V_{th}$ to left suggests that holes are trapped in trapping states and induce a positive gating ($\Delta V_g > 0$). Figure 4(d) is a schematic illustration of photogating. The Fermi level $E_F$ plus $\Delta V_g$ gives an increased new "effective" $E_F'$, which will trap more holes and enhance the photoresponse.[37] The photocurrent arising from photogating is proportional to $\Delta V_{th}$, i.e., the shift of $V_{th}$.[32] While the $I_{ph}$-$P$ relation is linear for the photoconductive effect, it is sublinear for photogating, i.e., $\alpha < 1$.[31,38] As shown above, the small values of $\alpha$ indicate that the dominant mechanism in BiSeI devices is photogating. The trapping states can be associated with the vacancies of coordination ions which induce doping levels in the bandgap. This has been confirmed in previous theoretical calculation results for BiSeI and similar semiconductors.[39, 40]

To conclude, the photodetectors are manufactured based on the exfoliated quasi-1D BiSeI thin wires of different thicknesses. Compared with the bulk device, a significantly enhanced photoresponse is observed. For the device on the 145 nm thick wire, the performance parameters are at least 3 orders of magnitude higher. The photoconductive effect is unlikely to account for the main mechanism, which should be understood in terms of the increased specific surface area and the photogating effect arising from the trapping states. This work sheds light on the modulation of optoelectronic properties and performance in quasi-1D materials.

**SUPPLEMENTARY MATERIAL**

See the supplementary material for more optoelectronic measurement data and analysis results.

This work was supported by the National Key R&D Program of China (Grant No. 2021YFA1600201), the National Natural Science Foundation of China (Grant Nos. 11874363, 11974356 and U1932216) and Anhui Province Laboratory of High Magnetic Field (Grant No. AHHM-FX-2020-01).

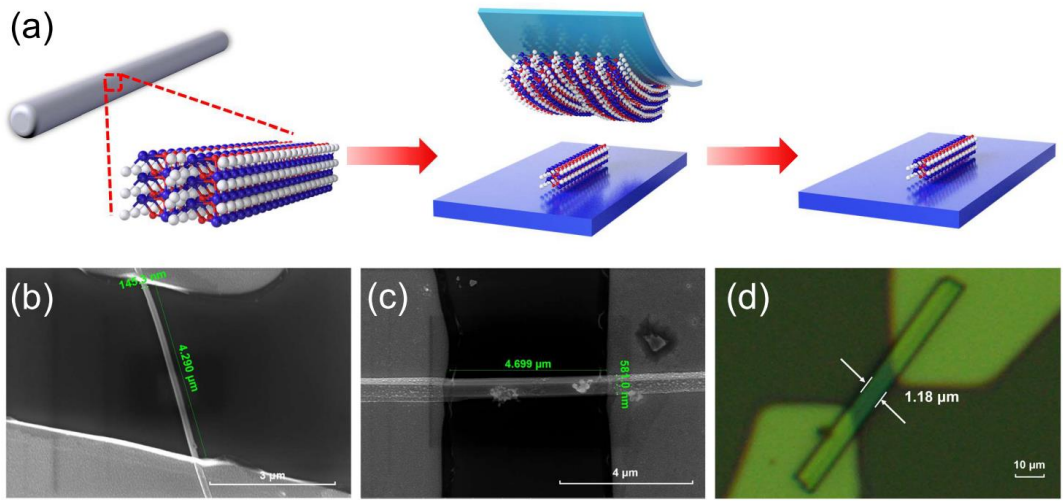

FIG. 1. (a) Schematic illustration of mechanical exfoliation and transfer process to obtain thin BiSeI wires. (b) SEM image of D1. (c) SEM image of D2. (d) Optical image of D3.

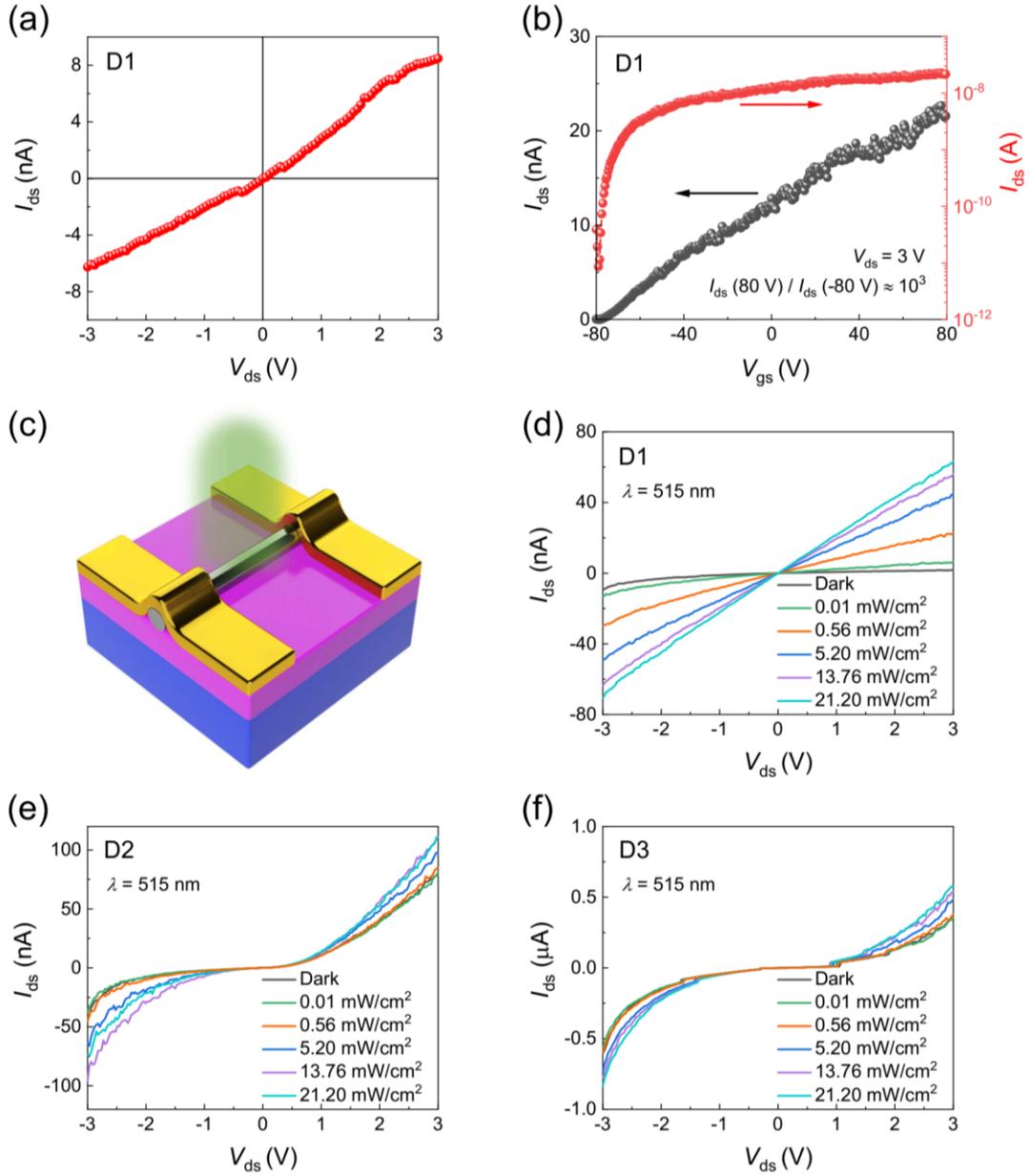

FIG. 2. (a) $I_{ds}$-$V_{ds}$ curve of D1. (b) Transfer curve of D1. (c) Schematic diagram of optoelectronic measurement. (d)-(f) $I_{ds}$-$V_{ds}$ curves of (d) D1, (e) D2 and (f) D3 taken in the dark and under illumination of a 515 nm laser tuned to different $P$.

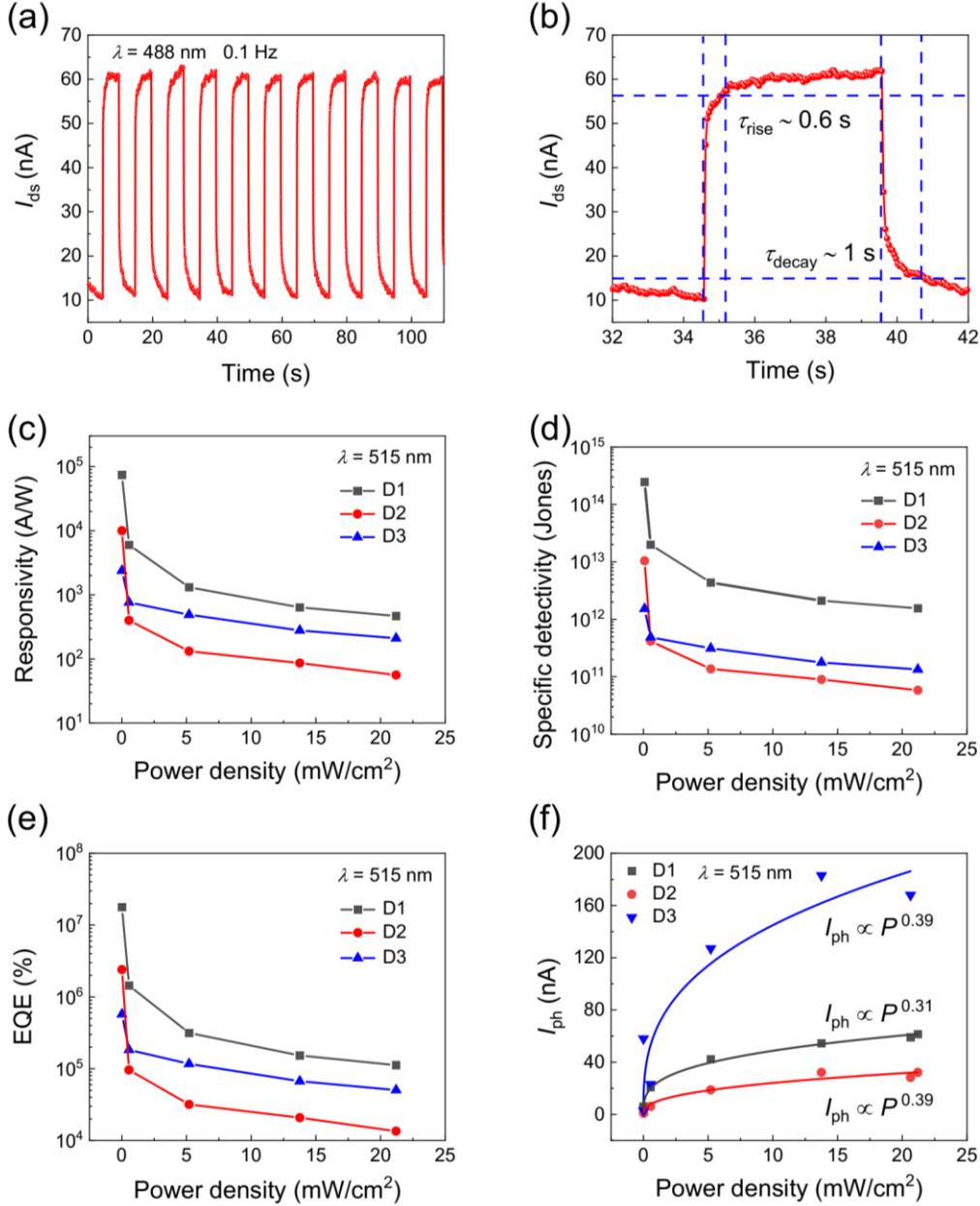

FIG. 3. (a) $I_{ds}$-$t$ curve of D1 taken at $V_{ds}$ = 3 V, $\lambda$ = 488 nm and $P$ = 21.2 mW cm$^{-2}$, showing a periodic ON/OFF response with a period of 10 s. (b) An enlarged view of a single response used to calculate response time. (c) Responsivity, (d) specific detectivity, (e) EQE and (f) photocurrent as a function of $P$ retrieved from the measurements taken at $V_{ds}$ = 3 V and $\lambda$ = 515 nm for D1, D2 and D3. The solid curves in (f) represent the power law fitting.

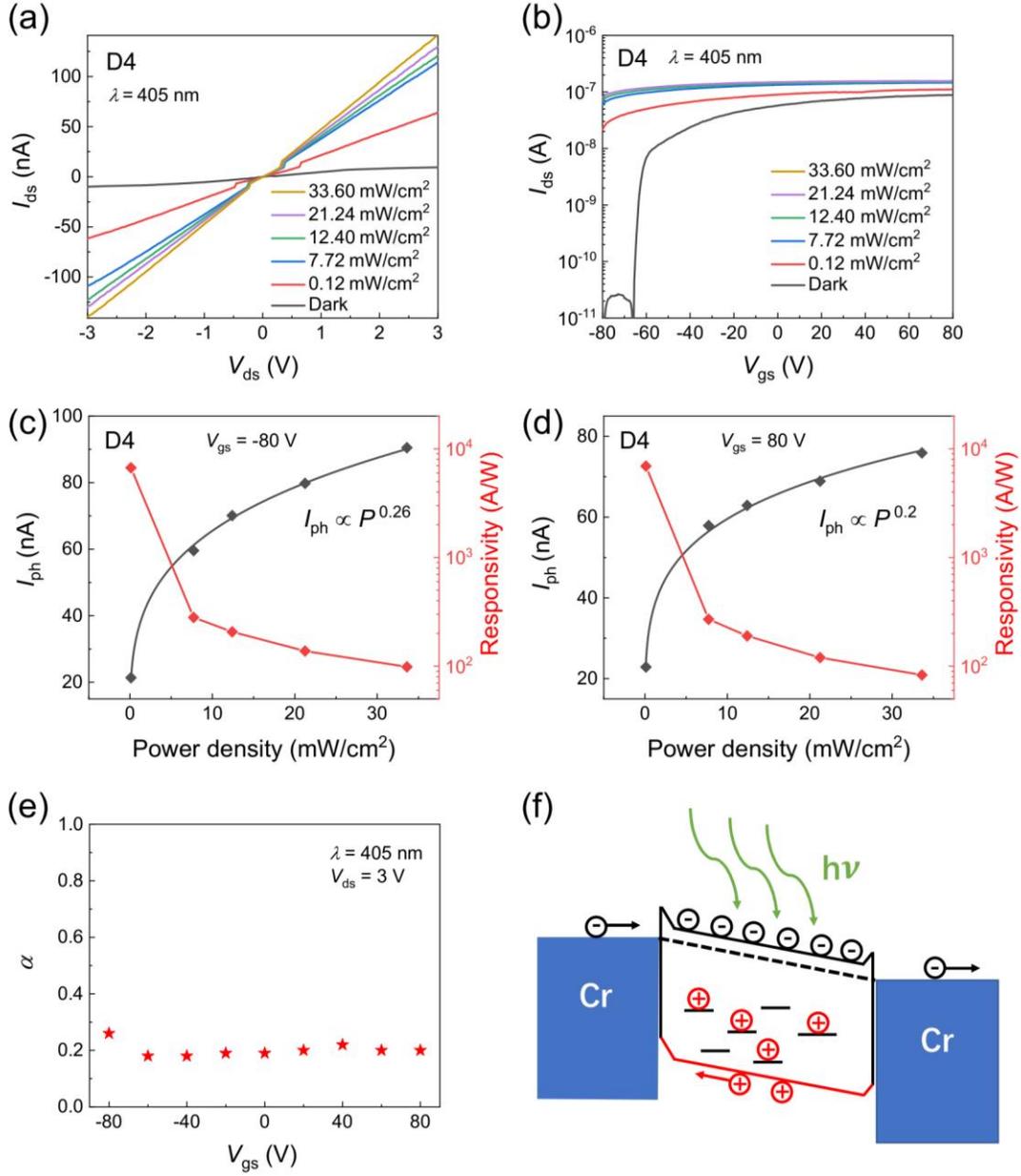

FIG. 4. (a) $I_{ds}$-$V_{ds}$ curves of D4 taken in the dark and under the 405 nm laser. (b) The semi-logarithmic scale plot of $I_{ds}$-$V_{gs}$ curves of D4 at $V_{ds}$ = 3 V. (c)(d) Photocurrent (left) and responsivity (right) as a function of $P$ retrieved from (a) for (c) $V_{gs}$ = -80 V and (d) $V_{gs}$ = 80 V. The power law fits of $I_{ph}$-$P$ are also plotted. (e) $\alpha$ as a function of $V_{gs}$. (f) Schematic energy bands to illustrate the photogating effect.

Summplementary material for

**Enhanced optoelectronic performance and photogating effect in quasi-one-dimensional BiSeI wires**

Hu et al.

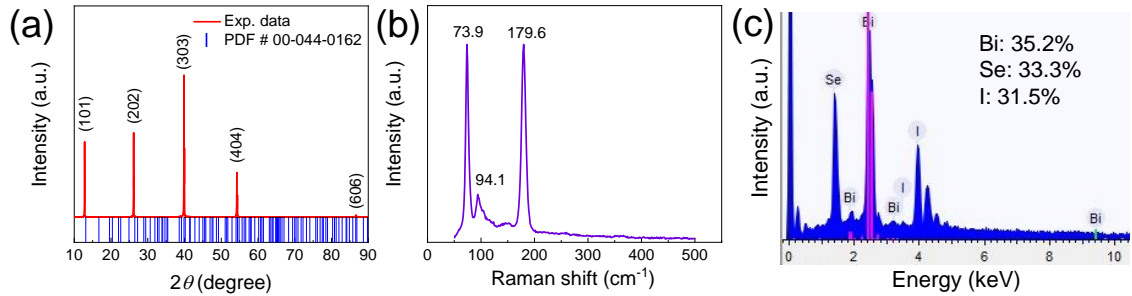

FIG. S1. (a) Single crystal XRD pattern. (b) Raman spectrum of a typical BiSeI wire. (c) EDS of BiSeI single crystal.

The structural and chemical characterizations have been performed for single crystal or thin wire samples. Figure S1(a) shows the XRD pattern acquired on a single crystal sample. Only a group of single crystal peaks are observed in the pattern, indicating that the sample is a single crystal and in pure phase. The diffraction peaks are indexed according to the powder XRD data (PDF No. 44-0162). Figure S1(b) shows the Raman spectrum of a typical BiSeI thin wire. The peak positions of the Raman vibration modes are consistent with the previous results obtained for bulk crystals,[1,2] indicating that the lattice vibrations in the thin-wire samples are almost the same as those of the bulk crystals. The chemical composition of BiSeI single crystal is further measured. As shown in Fig. S1(c), only the constituent elements of BiSeI are detected, and the atomic ratio is close to 1:1:1, again confirming the purity of the crystal. It is worth noting that there are some Se and I vacancies in the crystal that cannot be ignored. Their concentrations are calculated to be 5.4% and 10.5%, respectively.

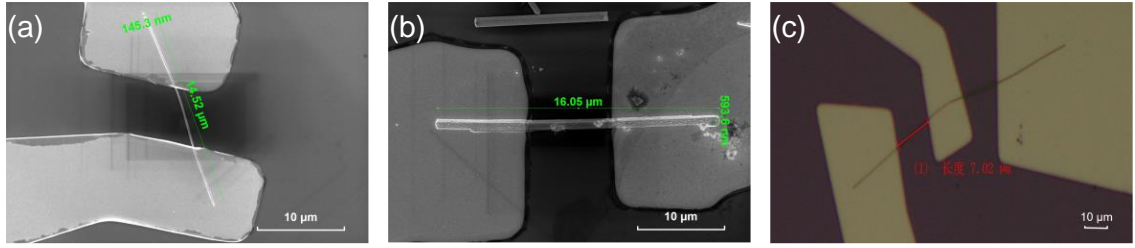

FIG. S2. (a) SEM image of D1. (b) SEM image of D2. (c) Optical image of D4.

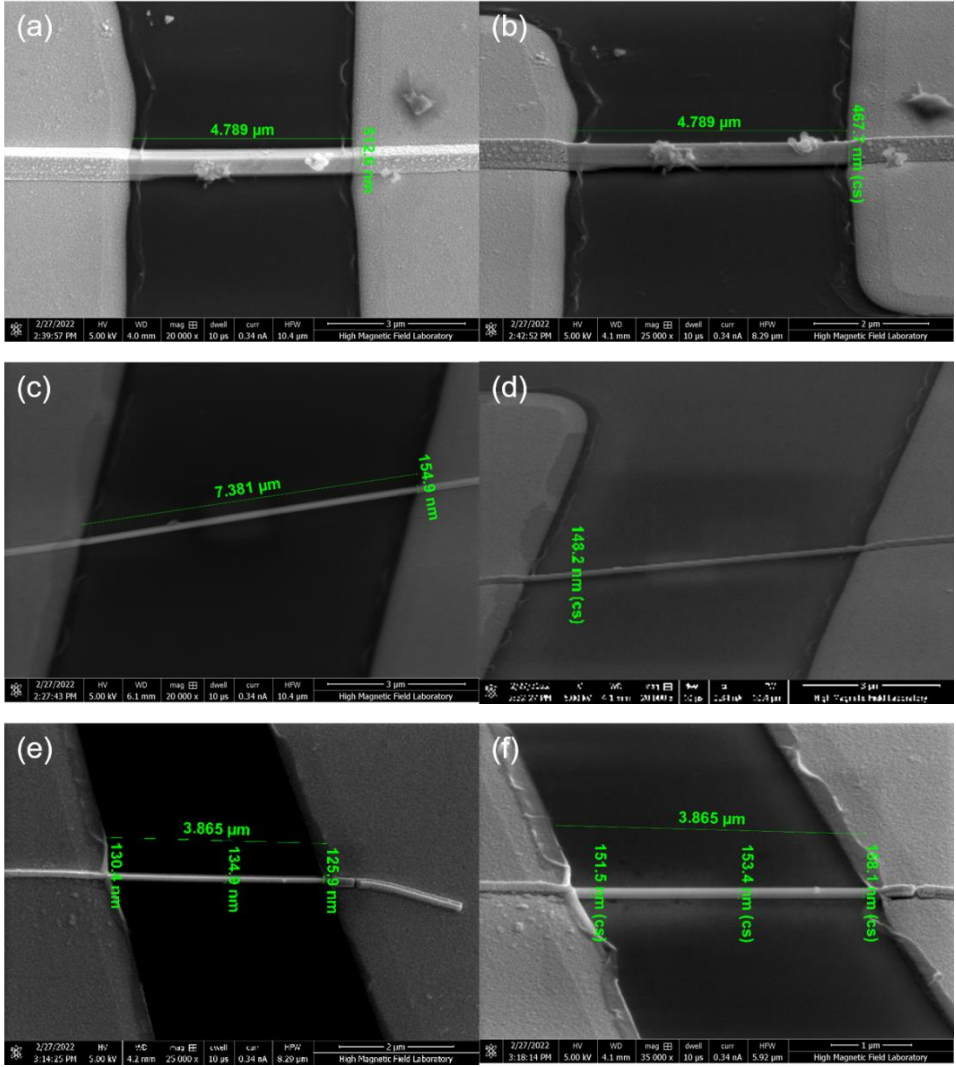

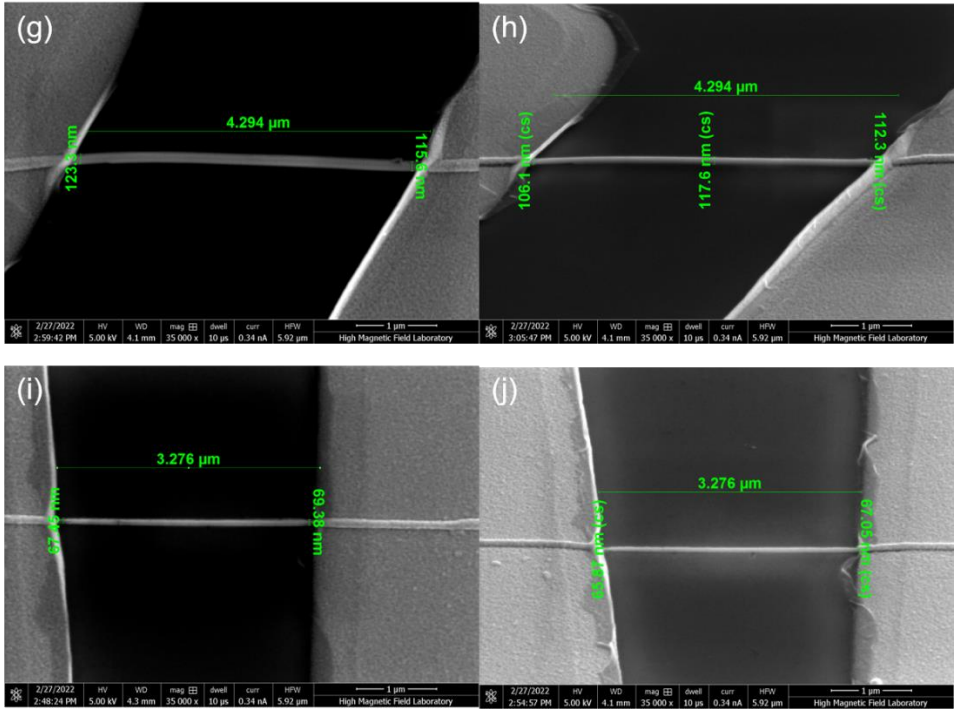

FIG. S3. SEM images of more BiSeI thin wires with different thicknesses. Each pair of images on the left and right panels represents the same sample. The images on the right panels were taken after rotating the sample holder 55° relative to the electron beam.

Since the electron beam has a remarkable impact on the BiSeI thin wires, the *I-V* curves cannot be repeated and the devices cannot be reused if observed under the electron beam. We suspect that absorption of electrons or loss of iodine ions may occur when the electron beam strikes the sample surface. Additional BiSeI wires have been measured to characterize their cross sections. Figure S3 shows SEM images of more BiSeI thin wires with different thicknesses. Each pair of images on the left and right panels represents the same sample. The images on the right panels were taken after rotating the sample holder 55° relative to the electron beam. We can find that as the thickness decreases, the width becomes more consistent for different viewing angles. That is, the wires become rounder. The roundness of thick wires is poor.

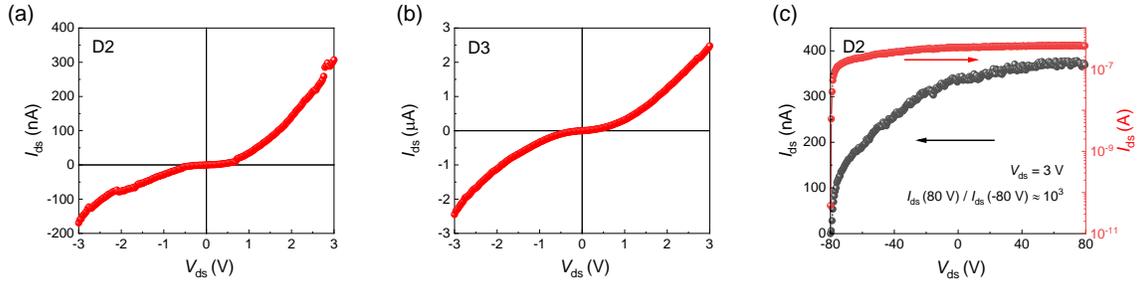

FIG. S4. (a) $I_{ds}$-$V_{ds}$ curve of D2. (b) $I_{ds}$-$V_{ds}$ curve of D3. (c) FET transfer curve ($I_{ds}$-$V_{gs}$) of D2.

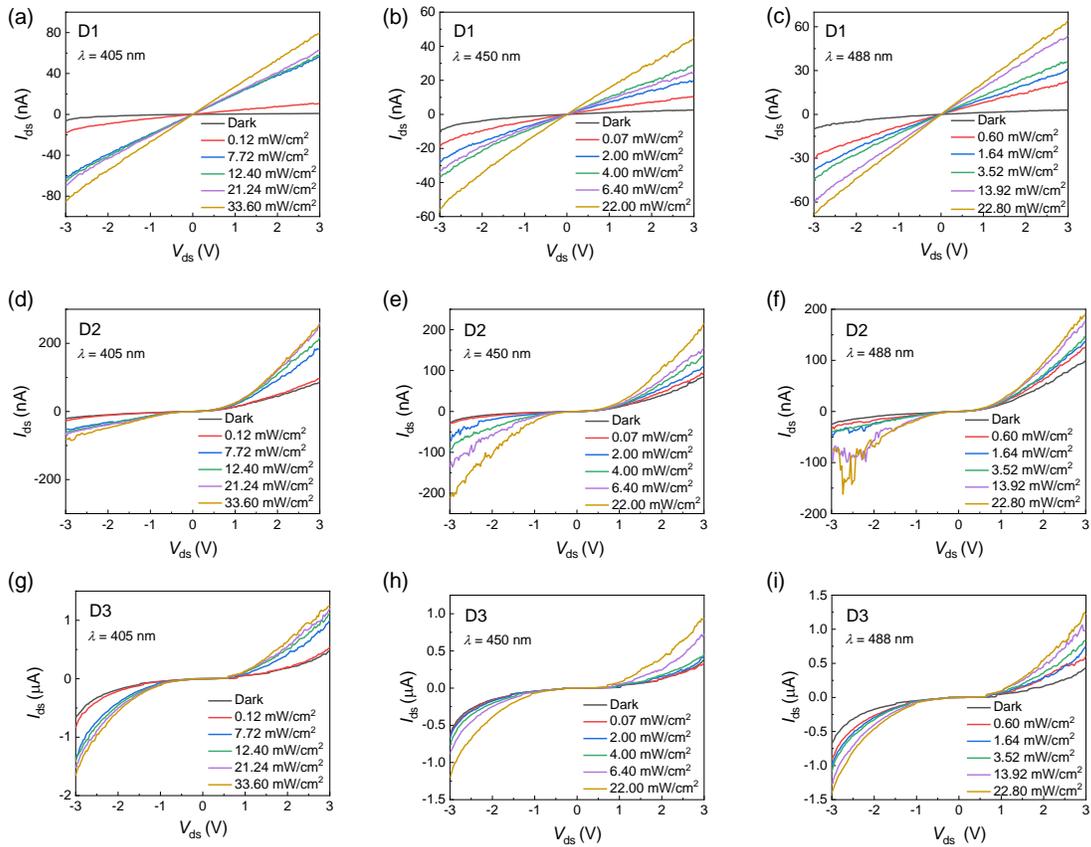

FIG. S5. $I_{ds}$-$V_{ds}$ curves of (a)-(c) D1, (d)-(f) D2 and (g)-(i) D3 taken in the dark and under illumination of a 405 nm laser, a 450 nm laser and a 488 nm laser tuned to different power densities.

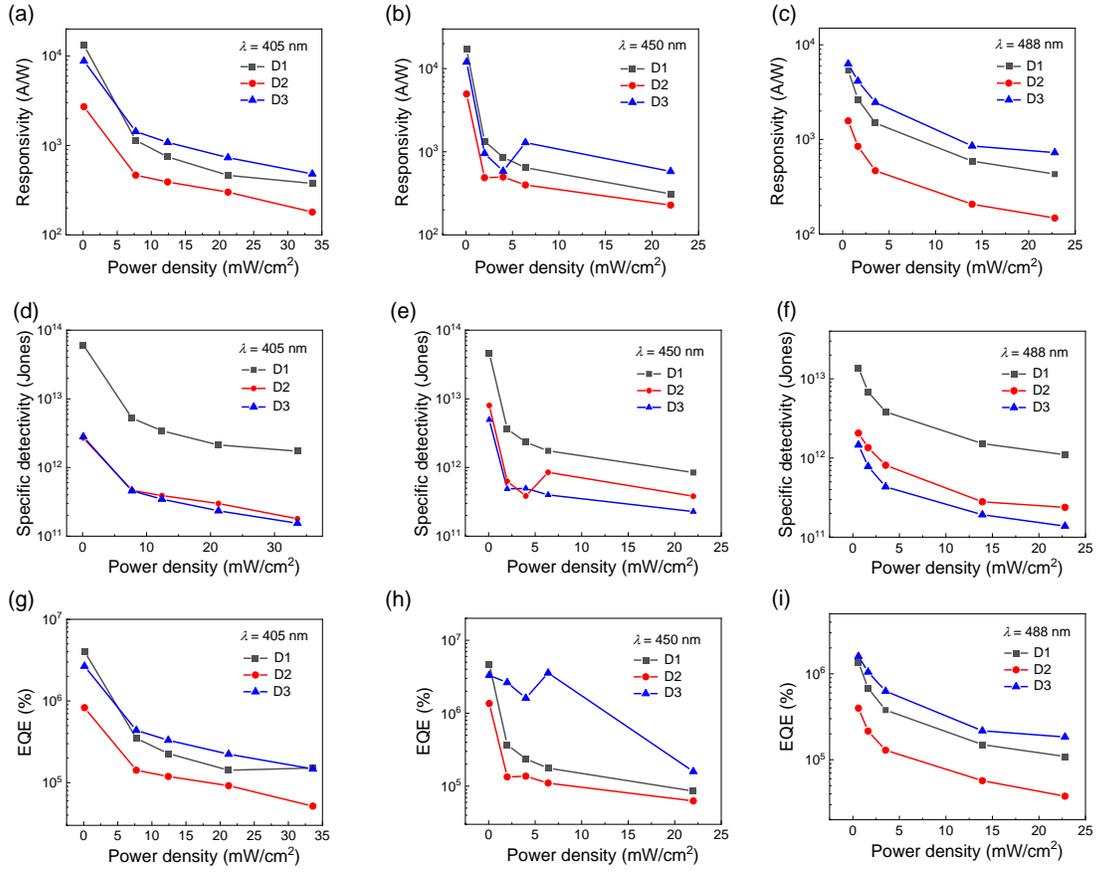

FIG. S6. (a)-(c) Responsivity, (d)-(f) specific detectivity and (g)-(i) external quantum efficiency as a function of laser power density retrieved from the optoelectronic measurements shown in Fig. S3. The calculations are taken at $V_{ds}$ = 3 V.

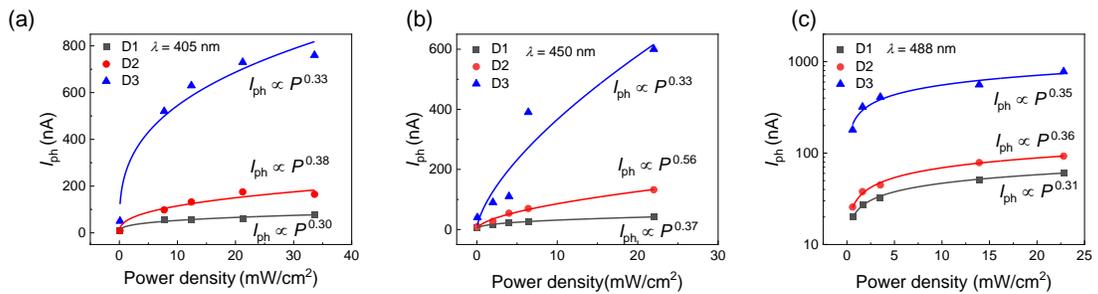

FIG. S7. Photocurrent as a function of laser power density retrieved from the optoelectronic measurements taken at $V_{ds}$ = 3 V shown in Fig. S3. The solid curves represent the power law fitting.

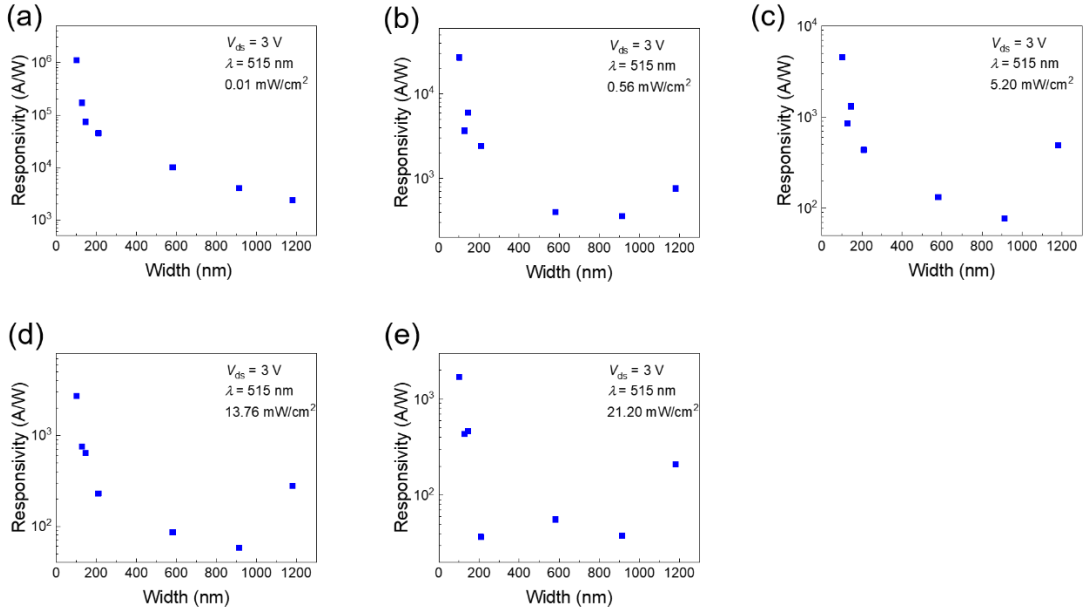

FIG. S8. Relationship between responsivity and channel width at different power densities. Additional devices are included.

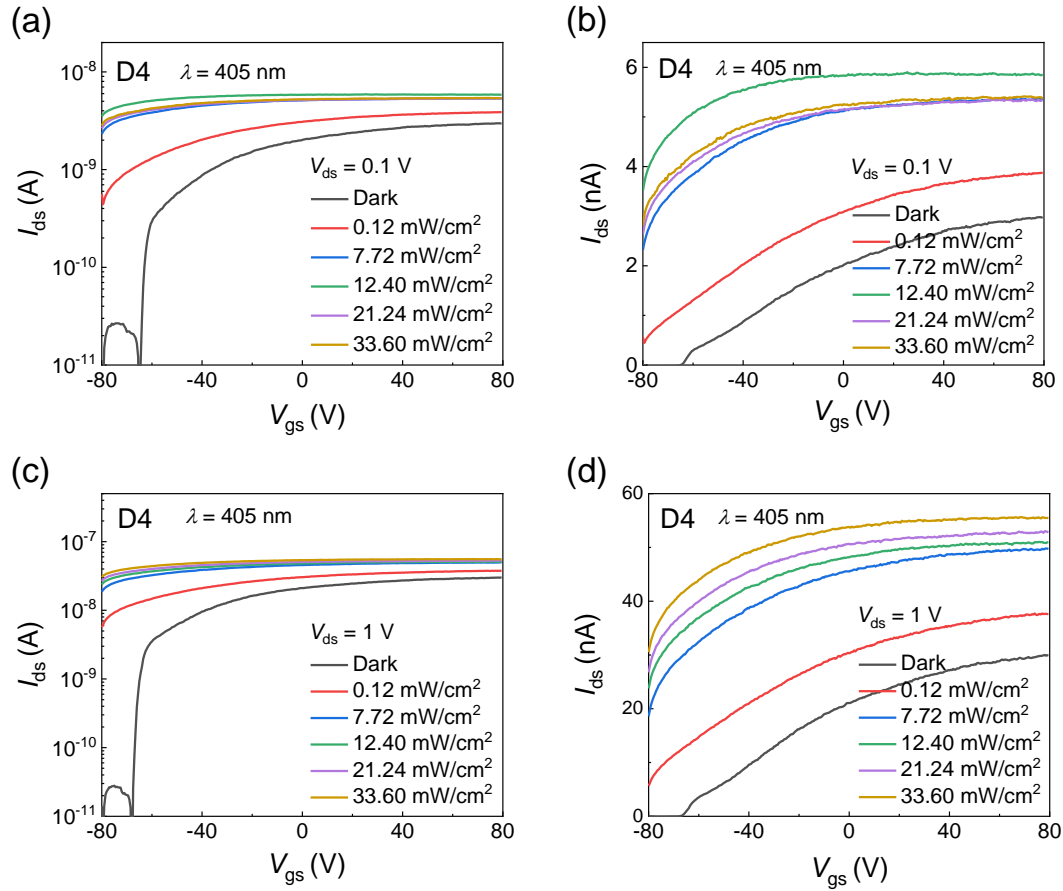

FIG. S9. $I_{ds}$-$V_{gs}$ curves of D4 taken at (a)(b) $V_{ds}$ = 0.1 V and (c)(d) $V_{ds}$ = 1 V in the dark and under the 405 nm laser tuned to different power densities.

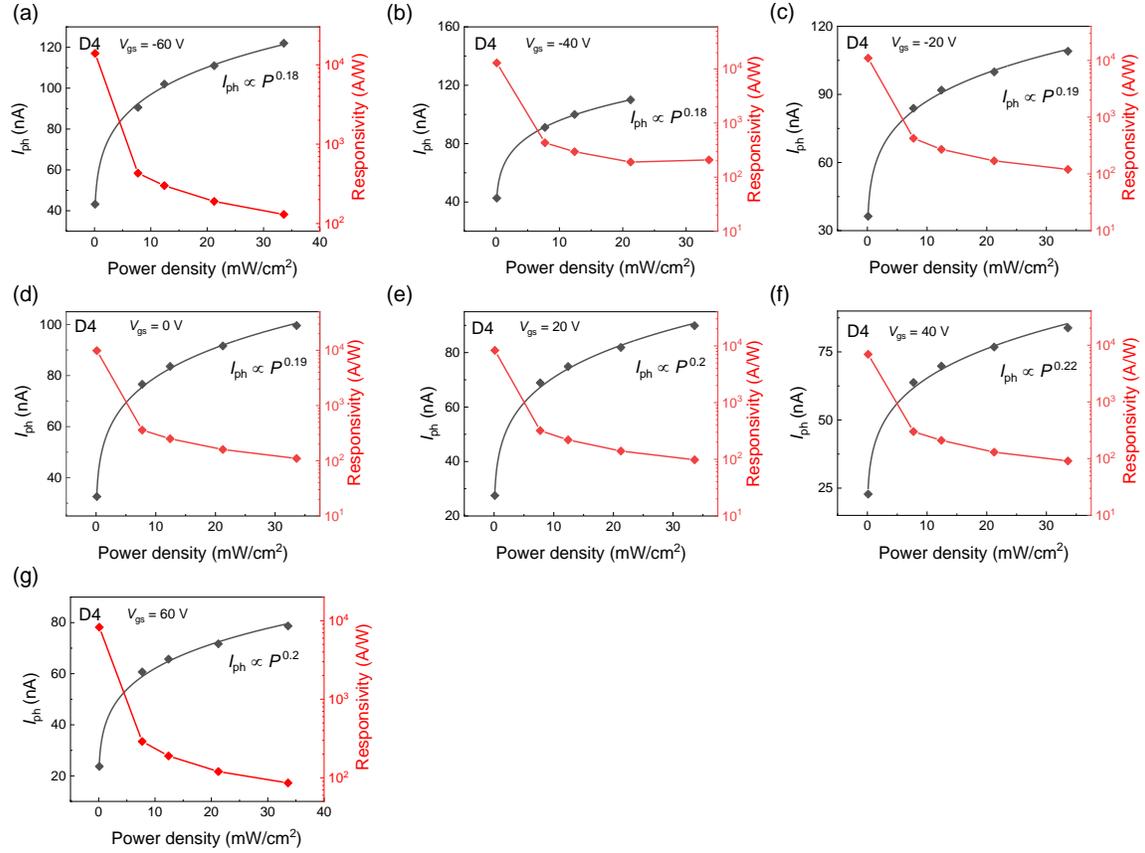

FIG. S10. Photocurrent (left) and responsivity (right) as a function of laser power density retrieved from Fig. 4(a) in the main text for $V_{gs}$ = -60 V, -40 V, -20 V, 0 V, 20 V, 40 V and 60 V.